\documentclass[twocolumn,preprintnumbers,floats,prd,amssymb,floatfix,nofootinbib,balancelastpage,superscriptaddress,amsmath]{revtex4-1}

\pdfoutput=1

\usepackage[utf8]{inputenc}
\usepackage[colorlinks=true,citecolor=blue,linkcolor=blue]{hyperref}
\usepackage[normalem]{ulem}
\usepackage{amsmath,amssymb,amstext}
\usepackage{epsfig}
\usepackage{graphicx}         
\usepackage{url} 
\usepackage{color}
\usepackage{slashed}
\usepackage{multirow}
\usepackage{placeins}
\usepackage[dvipsnames]{xcolor}
\usepackage{epstopdf}
\usepackage{soul} 
\usepackage{tikz}
\usepackage{mathtools}
\usepackage{xcolor}
\usepackage{multirow}

\begin{document}

\title{
Physics potential of a muon-proton collider
}

\author{Kingman Cheung}
\email{cheung@phys.nthu.edu.tw}
\affiliation{Department of Physics, Konkuk University, Seoul 05029, Republic of Korea}
\affiliation{Department of Physics, National Tsing Hua University, Hsinchu 300, Taiwan}

\author{Zeren Simon Wang}
\email{wzs@mx.nthu.edu.tw}
\affiliation{Department of Physics, National Tsing Hua University, Hsinchu 300, Taiwan}


%


\begin{abstract}

We propose a muon-proton collider with asymmetrical multi-TeV beam energies and integrated luminosities of $0.1-1$ ab$^{-1}$.
With its large center-of-mass energies and yet small Standard
Model background, such a machine can not only improve electroweak precision measurements but also probe new physics beyond the Standard Model to an unprecedented level.
We study its potential in measuring the Higgs properties, probing the R-parity-violating Supersymmetry, as well as testing heavy new physics in the muon $g-2$ anomaly.
We find that for these physics cases the muon-proton collider can perform better than both the ongoing and future high-energy collider experiments.

\end{abstract}

\maketitle


\section{Introduction}\label{sec:introduction}
The Large Hadron Collider (LHC) in Geneva, Switzerland has culminated with the discovery of a Standard Model (SM)-like Higgs boson in 2012 \cite{Aad:2012tfa,Chatrchyan:2012ufa}.
However, the world's largest machine has so far failed to find any new fundamental particles predicted by models beyond the Standard Model (BSM), e.g., the sleptons and squarks proposed by Supersymmetry (SUSY) models \cite{Nilles:1983ge,Martin:1997ns}. 
While the LHC program will be finished only in mid 2030's, a wide range of next-generation colliders have been proposed and intensively discussed.
For example, various $e^+ e^-$ colliders to run as a Higgs or $Z$-boson factory, have been suggested, including the International Linear Collider (ILC) \cite{Djouadi:2007ik,Behnke:2013xla}, Compact Linear Collider (CLIC) \cite{Charles:2018vfv}, the Future Circular Collider in the $ee$ mode (FCC-ee) \cite{Abada:2019zxq}, and the Circular Electron Positron Collider (CEPC) \cite{CEPCStudyGroup:2018ghi}.
Then, there are two main electron-proton collider proposals, i.e., the Large Hadron-electron Collider (LHeC) \cite{Klein:2009qt,AbelleiraFernandez:2012cc,Bruening:2013bga,Agostini:2020fmq} and FCC in the hadron-electron mode (FCC-he) \cite{Abada:2019lih,Benedikt:2018csr}, to be running concurrently with and after the HL-LHC, respectively.
In addition to these proposals, there are also higher-energy proton-proton colliders: High Energy LHC (HE-LHC) \cite{CidVidal:2018eel,Abada:2019ono,Azzi:2019yne}, FCC in the hadron-hadron mode (FCC-hh) \cite{Abada:2019lih,Benedikt:2018csr}, and Super Proton Proton Collider (SPPC) \cite{Tang:2015qga}, as well as
renewed interests in muon colliders \cite{Han:2012rb,Delahaye:2019omf,Costantini:2020stv,Han:2020uak,Buttazzo:2020eyl,Yin:2020afe,Capdevilla:2020qel,Han:2020pif,Long:2020wfp,Huang:2021nkl,Capdevilla:2021rwo,Liu:2021jyc,Asadi:2021gah}\footnote{More recently electron-muon head-on collisons have also been studied for the first time in Refs.~\cite{Lu:2020dkx,Bossi:2020yne}.}.
The experiments in these colliders would for instance excel in electroweak (EW) precision measurements, Quantum Chromodynamics (QCD) tests, or BSM physics searches.

In this work, we consider a muon-proton collider with multi-TeV beam energies.
Historically this type of machine was first investigated in late 1990's \cite{Shiltsev:1997pv,Cheung:1999wy,Carena:2000su}, and more recently in Refs.~\cite{Caliskan:2017meb,Acar:2017eli}, focusing either on a setup of a center-of-mass (CM) energy of several hundred GeV only or on compositeness models.
Compared to the other types of colliders, a $\mu p$ collider has its unique advantages.
First, while synchrotron radiation prevents a circular electron beam from obtaining high energies, this issue is much more tamed for a muon beam, allowing a $\mu p$ collider to achieve a much higher center-of-mass energy and thus much larger scattering cross sections in general than an electron-proton collider.
Further, a $\mu p$ collider shares the upside of an $e p$ collider such that BSM studies on this type of machine5s usually suffer from smaller QCD backgrounds, than at $pp$ collisions.
Moreover, with multi-TeV CM energies, a $\mu p$ collider could produce TeV-scale new particles on shell, which is, however, more difficult to achieve at, e.g., a multi-TeV muon collider.

There are admittedly downsides of a $\mu p$ collider.
Notably muons are short-lived.
This requires a sufficiently large acceleration for the muon beam so that the muons reach the interaction point before decaying, and a careful examination of the beam-induced background (BIB).
As Ref.~\cite{Lu:2020dkx} pointed out, the BIB can be reduced by a large extent if the signal final-state particles are largely boosted towards the other beam side\footnote{This strategy is, however, inapplicable for muon colliders which suffer from the BIB issue on both sides of the beams.}.
As we will see, because of the proton parton distribution, this is indeed the case for $\mu p$ collisions even if the proton beam energy is one order of magnitude larger than the muon one.

Given the discussion of $\mu p$ collisions above, one can easily see that multi-TeV $\mu p$ colliders can probe a much higher scale in
  deep-inelastic scattering than other collider experiments such as the LHeC.
  For instance, with a CM energy of 5 TeV, the largest potential reach in
  momentum squared transfer, $Q^2$, can be of order $10^7$ GeV$^{2}$.
In the present paper, however, we will focus on studying the potential of $\mu p$ colliders in probing BSM physics.

The organization of this work is as follows.
In Sec.~II we introduce the relevant parameters of the two tentative $\mu p$ collider setups we propose.
We then study in detail in Secs.~III, IV, and V, the sensitivity
reach of these potential experiments in Higgs coupling measurements, R-parity-violating Supersymmetry, and finally heavy new physics (NP) in the muon $g-2$.
We summarize in Sec.~V.

\section{Collider setups}\label{sec:Collidersetups}
In this work we focus on two possible beam combinations: (1) ``$\mu p-1$'' with $E_p=7$ TeV and $E_{\mu^-}=1$ TeV, and (2) ``$\mu p-2$'' with $E_p=50$ TeV and $E_{\mu^-}=3$ TeV.
The proton beam energies are in agreement with the HL-LHC and FCC, while the muon energies are inspired from the current discussion on TeV-scale muon colliders.

Estimates on the instantaneous luminosity at muon-proton colliders
where performed in the past \cite{Shiltsev:1997pv,Ginzburg:1998yw,Kaya:2019ecf}.
In general, realistic estimates for the luminosity given the current technologies should be at the order of $10^{33}$ cm$^{-2}$ s$^{-1}$, which we assume for $\mu p-1$.
As for $\mu p-2$ which is supposed to be an upgrade of $\mu p-1$, we take a slightly optimistic value of $10^{34}$ cm$^{-2}$ s$^{-1}$.
For the lifespan of these experiments, we take as a benchmark operation time $10^{7}$ s/year for $10$ years, leading to an integrated luminosity $\mathcal{L}^{\text{int}}$ of $0.1$ ab$^{-1}$ and 1 ab$^{-1}$ for $\mu p-1$ and $\mu p-2$, respectively.
We summarize these collider parameters in Table~\ref{tab:experiments}.
\begin{table}[t]
	\begin{center}
		\begin{tabular}{c|c|c|c|c}
			\hline
			Exp.	& $E_p~[\text{TeV}]$  & $E_{\mu^-}~[\text{TeV}]$  & $\sqrt{s}~[\text{TeV}]$ & $\mathcal{L}^{\text{int}}~[\text{ab}^{-1}]$\\
			\hline
			$\mu p-1$ &  7  & 1 & 5.3 &  0.1\\
			$\mu p-2$ &  50 & 3 & 24.5  &   1\\
			\hline 
		\end{tabular}
		\caption{
			Basic parameters of the two $\mu p$ experiments considered in this work.
		}
		\label{tab:experiments}
	\end{center}
\end{table}

\section{Higgs precision measurements}\label{sec:Higgs}
One of the utmost tasks in Higgs physics is the precision
measurements of the Higgs boson couplings.
Here we study the projected uncertainties in the measurement of the Higgs coupling to $b$-quarks at a $\mu p$ collider.

\begin{figure}[t]
	\includegraphics[width=0.45\textwidth]{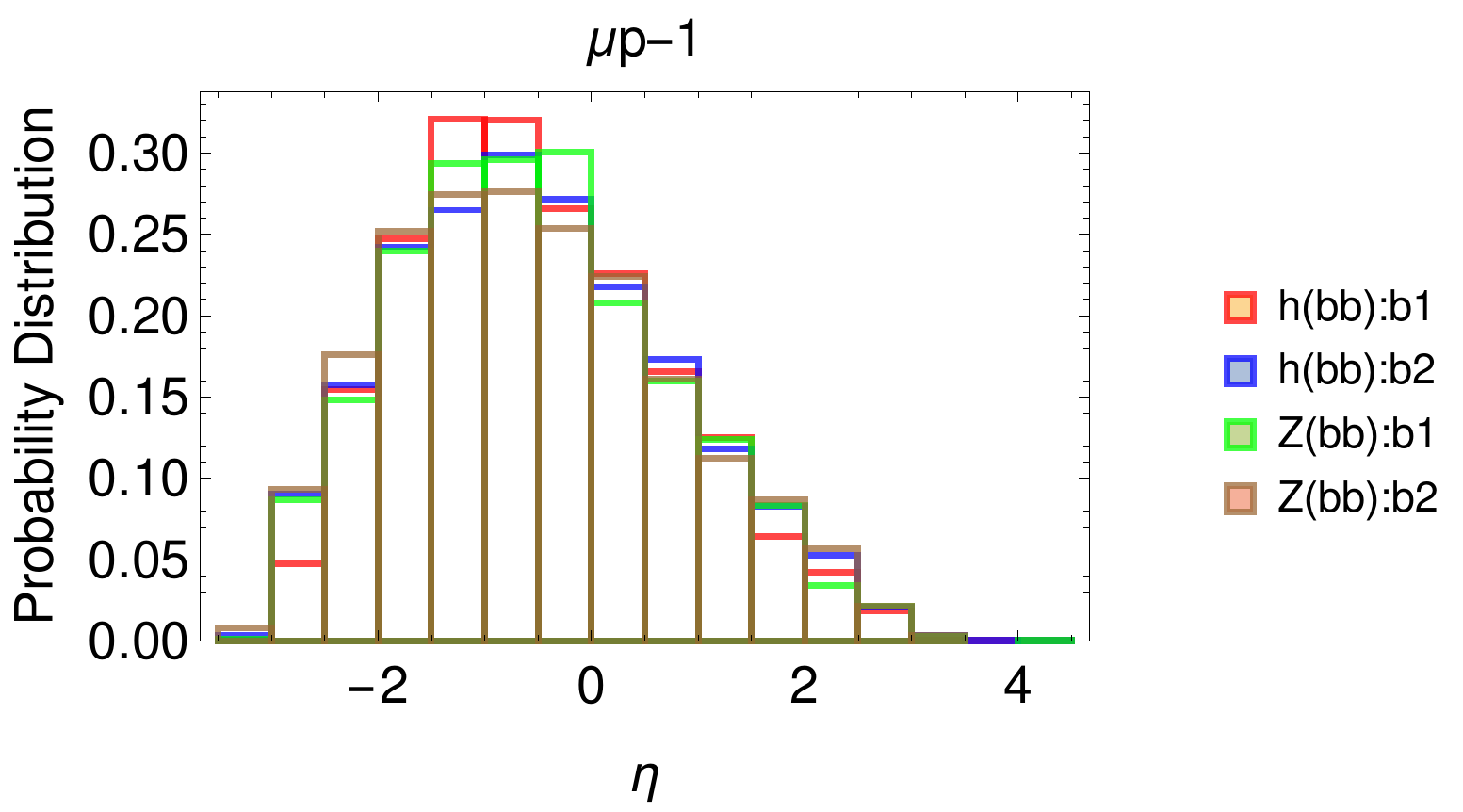}\\
	\vspace{0.3cm}
	\includegraphics[width=0.45\textwidth]{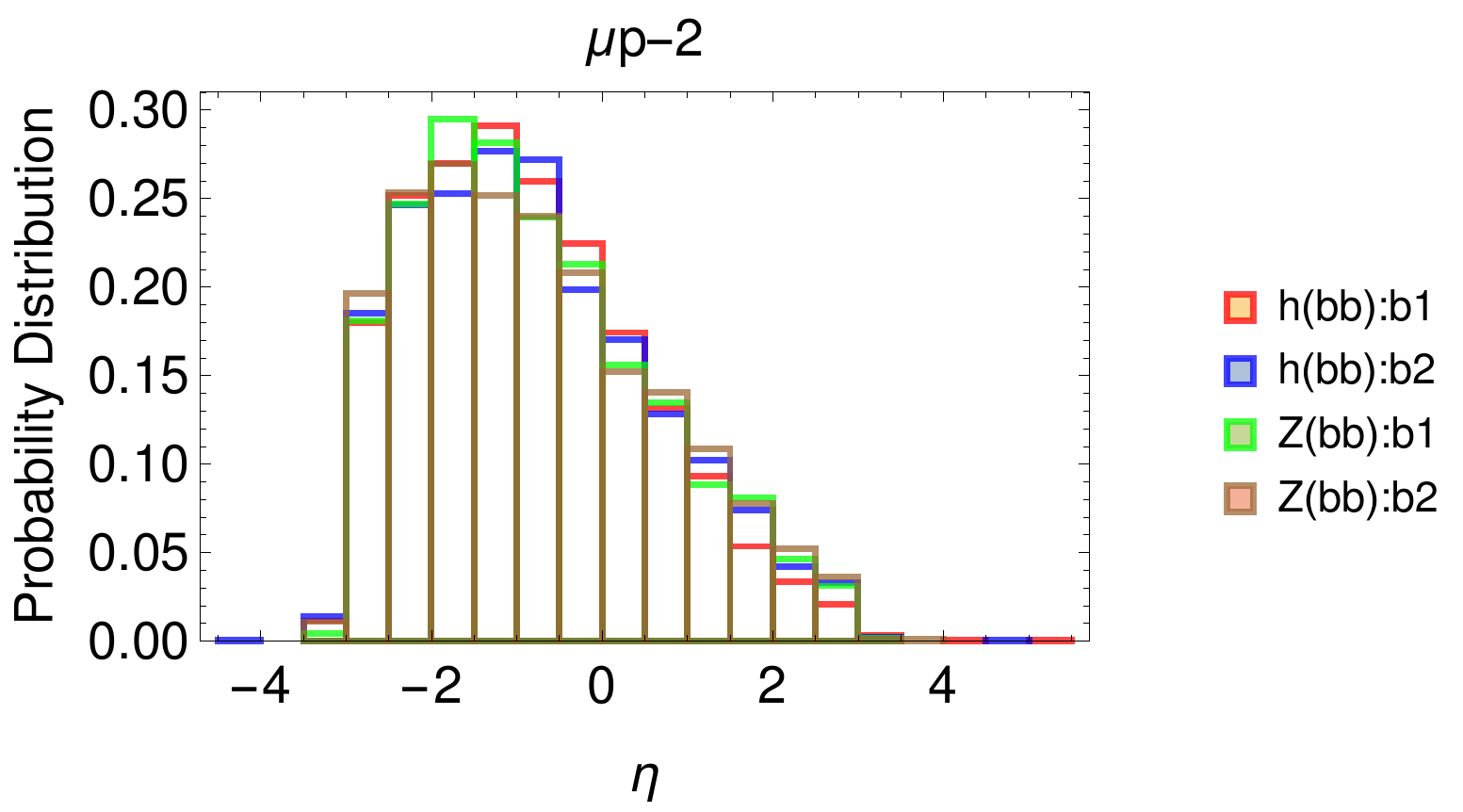}
	\caption{Pseudorapidity distributions of the reconstructed $b$-jets for the signal and background events at $\mu p-1$ and $\mu p-2$.
	}
	\label{fig:eta_b}
\end{figure}

Similar to $ep$ collisions, the Higgs boson at $\mu p$ is produced mainly via the  $WW$ and $ZZ$ vector-boson-fusion (VBF) processes.
\begin{table}[t]
	\begin{tabular}{c|c|c|c|c|c}
		\hline
		\multicolumn{1}{c|}{VBF process} & $\mu p-1$ & $\mu p-2$ & LHeC & FCC-he & LHC-14        \\ \hline
		$WW$                              &       0.978    & 5.103          & 0.110 \cite{Agostini:2020fmq}    &  0.577 \cite{Agostini:2020fmq}    & \multirow{2}{*}{4.233 \cite{lhcHiggsXS}}  \\
		$ZZ$                              &0.216           &1.263           &  0.020 \cite{Agostini:2020fmq}   &   0.127  \cite{Agostini:2020fmq}     &                   \\ \hline
	\end{tabular}
	\caption{The inclusive cross sections of VBF production of the SM Higgs bosons at various experiments, in pb. }
	\label{tab:VBFXS}
\end{table}
In Table~\ref{tab:VBFXS} we list the \textit{inclusive} production cross sections of the SM Higgs boson at $\mu p-1$, $\mu p-2$, LHeC, FCC-he, and the LHC with $\sqrt{s}=14$ TeV.
We find that the VBF cross sections at $\mu p-1$ and $\mu p-2$, obtained at leading order with MadGraph5 3.0.2 \cite{Alwall:2014hca}, can be up to about one order of magnitude larger than those at the LHeC and FCC-he, and even comparable to those at the LHC with $\sqrt{s}=14$ TeV.
Here, we choose to focus on the $WW$ VBF process: $p \mu^- \to j \nu_\mu h, h \to b \bar{b}$ because of its larger rate than the $ZZ$ process.
The dominant background is the corresponding $WW$ VBF for
$Z$-boson production with $Z\to b\bar{b}$.
We express the measurement uncertainty of the cross section of $p \mu^-\to j \nu_\mu b \bar{b}$ as $\Delta\sigma/\sigma=\sqrt{N_s+N_b}/(N_s)$ including the statistical error only, where $N_{s/b}$ denotes the signal/background event numbers, and perform a cut-based analysis to estimate the sensitivity reach in $\Delta\sigma/\sigma$.
\begin{table}[t]
	\begin{center}
		\begin{tabular}{c|c|c|c|c|c|c|c}
			\hline
			Exp. &  $ \epsilon_{hbb}^{\text{pr-cut}}$ &$ \sigma(p \mu^- \to j \nu_\mu Z)$ & $ \epsilon_{zbb}^{\text{pr-cut}}$ & $\epsilon_\text{cut}^{\text{sig}}$& $\epsilon_\text{cut}^{\text{bgd}}$ & $\frac{\Delta \sigma}{\sigma}$& $\frac{\Delta g_{hbb}}{g_{hbb}}$\\
			\hline
			$\mu p-1$ & $0.98$ & $4.67$ pb & $0.99$ & $0.21$ &  $0.022$ & $0.97\%$& $0.69\%$\\
			$\mu p-2$ &  $0.91$ & $25.1$ pb  & $0.97$ & $0.17$ &  $0.018$ & $0.15\%$ & $0.50\%$ \\
			\hline 
		\end{tabular}
		\caption{
			Summary of parton-level and reconstructed-level cut efficiencies, and inclusive production cross section of the background process at $\mu p-1$ and $\mu p-2$.
		    The last columns list the final reaches in $\Delta\sigma/\sigma$ and $\Delta g_{hbb}/g_{hbb}$.
		}
		\label{tab:hbb_results}
	\end{center}
\end{table}
We generate the parton-level events with MadGraph5, requiring $p_T^{j/b}>5$ GeV and $|\eta^{j/b}|<5.5$.
The $p_T$ threshold avoids the collinear limit, and the $|\eta^{j/b}|$ cut corresponds to the geometry of the beam-asymmetric LHeC detector.
The parton showering and hadronization for asymmetrical lepton-hadron collisions are properly treated with a patched version of Pythia 6.428 \cite{Sjostrand:2006za, py6_patched}.
Finally we perform jet clustering with FastJet 3.3.2 \cite{Cacciari:2011ma,Cacciari:2005hq} with the anti-$k_t$ algorithm \cite{Cacciari:2008gp}, and fast detector simulation with Delphes 3.4.2 \cite{deFavereau:2013fsa}.
For the latter we use an LHeC-specific Delphes card which includes the beam asymmetry.
For $b$-tagging efficiency we take $75\%$.
The following set of cuts at the reconstructed level are imposed.
We first keep only the events with exactly two $b$-jets.
In Fig.~\ref{fig:eta_b} we show the pseudorapidity distributions of the
$b$-jets.
We find the produced $b$'s are peaked at the proton beam side due to the proton parton distributions and expected to allow for BIB reduction.
We then select only events where the $b$-jet pair invariant mass, $m_{bb}$,
is close to the Higgs mass 125 GeV: $|m_{bb}-m_h|<25$ GeV, intended to separate the signal and background events.
After these event selections, we compute the signal and background event numbers with $N_{s}=\mathcal{L}^{\text{int}} \cdot \sigma(p \mu^- \to j \nu_\mu h) \cdot \text{Br}( h \to b \bar{b} ) \cdot \epsilon_{hbb}^{\text{pr-cut}}\cdot \epsilon_\text{cut}^{\text{sig}}$ and $N_{b}=\mathcal{L}^{\text{int}} \cdot \sigma(p \mu^- \to j \nu_\mu Z)\cdot \text{Br}( Z \to b \bar{b})\cdot \epsilon_{Zbb}^{\text{pr-cut}}\cdot \epsilon_\text{cut}^{\text{bgd}}$.
$\sigma(p \mu^- \to j \nu_\mu h)$ was given in Table~\ref{tab:VBFXS}, $\epsilon_{hbb}^{\text{pr-cut}}$ and $\epsilon_{Zbb}^{\text{pr-cut}}$ measure the reduction on the signal and background production cross sections from the parton-level cuts, and $\epsilon_\text{cut}^{\text{sig}}$ and $\epsilon_\text{cut}^{\text{bgd}}$ are the reconstructed-level cut efficiencies.
These are all listed in Table~\ref{tab:hbb_results} together with the inclusive cross sections $\sigma(p \mu^- \to j \nu_\mu Z)$.
The projected reaches in $\Delta\sigma/\sigma$ at $\mu p-1$ and $\mu p-2$ are thus estimated as $0.97\%$ and $0.15\%$, respectively.
In order to translate the uncertainties on $\sigma$ to those on the Higgs-$b$-$\bar{b}$ coupling, $g_{hbb}$, we need to take into account the measurement uncertainties on $\frac{g^2_{hWW}}{\Gamma_h}$ where $g_{hWW}$ and $\Gamma_h$ are the Higgs coupling to the $W$-bosons and the Higgs total decay width.
$\Delta \big( \frac{g^2_{hWW}}{\Gamma_h} \big)/\frac{g^2_{hWW}}{\Gamma_h}$ can be derived from the uncertainties on the Higgsstrahlung production cross section and its product with Br($h\to WW$) at the FCC-ee as benchmark values: $\sigma(e^- e^+ \to Z h)$ and $\sigma(e^- e^+ \to Z h)\cdot$Br$(h\to WW)$.
These have been given as 0.4\% and 0.9\% in e.g., Ref.~\cite{Ruan:2014xxa}, allowing us to estimate $\Delta \big( \frac{g^2_{hWW}}{\Gamma_h} \big)/\frac{g^2_{hWW}}{\Gamma_h}$ as $\sqrt{(0.4\%)^2+(0.9\%)^2}=0.985\%$, as $\sigma(e^- e^+ \to Z h)\cdot$Br$(h\to WW)/\sigma(e^- e^+ \to Z h)=g^2_{hWW}/\Gamma_h$.
Since $\sigma(\mu^- p \to \nu_\mu j b \bar{b})=\sigma(\mu^- p \to \nu_\mu j h)\cdot \text{Br}(h\to b\bar{b})=\frac{g^2_{hWW}g^2_{hbb}}{\Gamma_h}$, the uncertainty on $g_{hbb}$ can be computed with $\frac{\Delta g_{hbb}}{g_{hbb}}=\frac{1}{2}\sqrt{\big(\frac{\Delta\sigma}{\sigma}\big)^2+\Bigg(  \frac{\Delta \big( \frac{g^2_{hWW}}{\Gamma_h} \big)}{\frac{g^2_{hWW}}{\Gamma_h}}   \Bigg)^2}$, which leads to 0.69\% and 0.50\% for $\mu p-1$ and $\mu p-2$, respectively, in comparison with $0.97\%$ at the LHeC obtained by a cut-based analysis \cite{lhecKappahbb}, and $4\%$ at the CMS experiment with 3 ab$^{-1}$ integrated luminosity \cite{CMS:2013xfa,Cepeda:2019klc}.

We comment that a similar improvement in measuring the other Higgs couplings such as those to the gauge bosons is also expected.

\section{R-parity-violating Supersymmetry}\label{sec:rpv}
\begin{figure}[t]
	\includegraphics[width=0.45\textwidth]{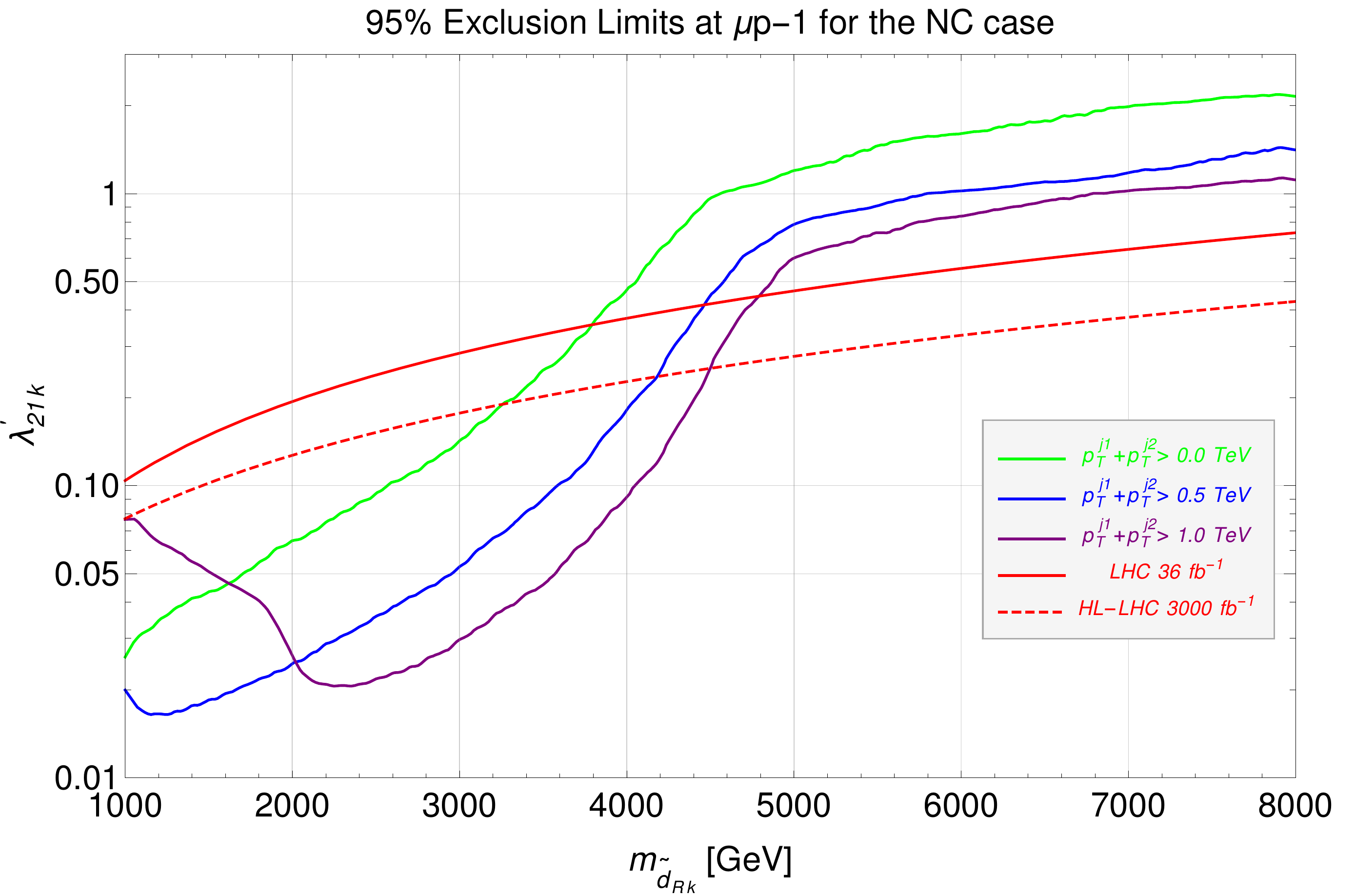}
	\includegraphics[width=0.45\textwidth]{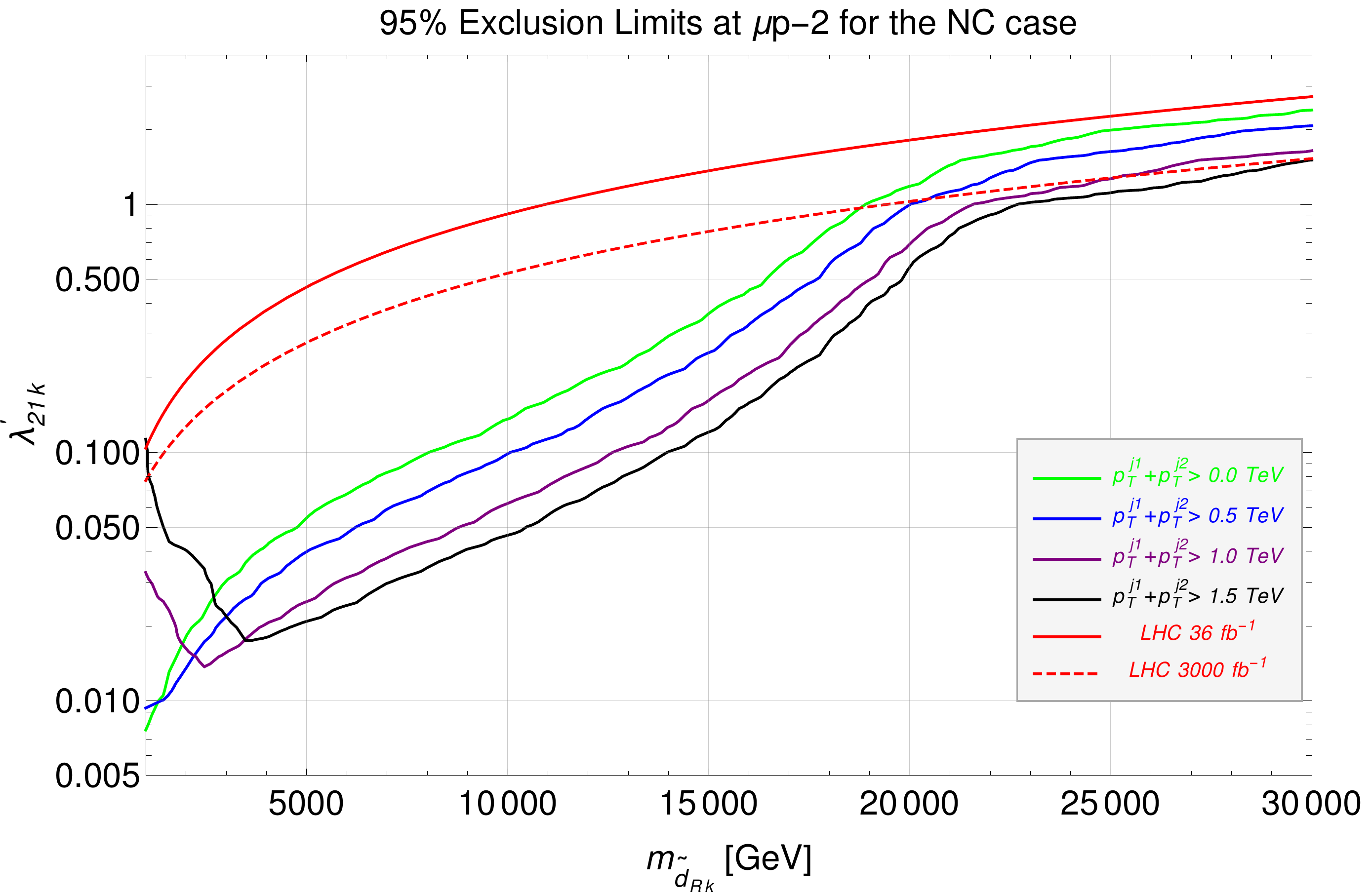}
	\caption{$95\%$ C.L. exclusion limits for $\lambda'_{21k}$ vs. $m_{\tilde{d}_{Rk}}$.
	}
	\label{fig:rpv_nc}
\end{figure}

Even though no new particles havee been discovered at the LHC and
TeV-scale lower mass bounds on the squarks and gluinos have been established
\cite{Aaboud:2018doq,Sirunyan:2017nyt,Sirunyan:2019mbp,Sirunyan:2019ctn,Aad:2020nyj}, SUSY remains one of the most motivated BSM models.
In SUSY, a $Z_2$ parity, known as R-parity, is usually assumed, rendering the proton stable and offering the lightest supersymmetric particle as a dark matter candidate.
However, it is equally legitimate to consider the R-parity-violating Supersymmetry (RPV-SUSY) scenario (see Refs.~\cite{Dreiner:1997uz,Barbier:2004ez,Mohapatra:2015fua} for reviews).
The latter, in fact, offers rich phenomenology at colliders.
With the broken R-parity, the superpotential of the Minimal Supersymmetric Standard Model (MSSM) is extended with:
\begin{eqnarray}
	W_{\text{RPV}} = &&\epsilon_i L_i \cdot H_u + \frac{1}{2} \lambda_{ijk} L_i \cdot L_j \bar E_k + \lambda'_{ijk} L_i \cdot Q_j \bar D_k \nonumber\\
	&& + 
	\frac{1}{2}\lambda''_{ijk}\bar U_i \bar D_j \bar D_k, \label{eq:RPVsuperpotential}
\end{eqnarray}
where the operators in the first line violate lepton numbers and those in the second line violate baryon numbers.
Allowing all these terms to be non-vanishing would lead to a too fast proton decay rate unless the couplings are extremely small.
For the purpose of this work, we focus on the operator $\lambda'_{ijk}L_i\cdot Q_j \bar{D}_k$ while assuming the others are vanishing\footnote{This can be justified by e.g., imposing a baryon triality $B_3$ symmetry \cite{Ibanez:1991pr,Dreiner:2012ae}.}.
In particular, here the RPV squark is a specific leptoquark which was used to explain a number of flavor anomalies \cite{ColuccioLeskow:2016dox,Crivellin:2020mjs,Crivellin:2020tsz}.
Ref.~\cite{Carena:2000su} from two decades ago performed an analytic estimate of sensitivity reach at a high-energy muon-proton collider (with $E_{\mu^{\pm}}=200$ GeV and $E_p=1$ TeV) to the RPV couplings $\lambda'_{2j1}$ and $\lambda'_{21k}$ for squark masses below 1 TeV.
In this work, we focus on one Drell-Yan-like signal process as an example: $p \mu^- \to \mu^- u$ (neutral current, or denoted as ``NC''), mediated by a right-chiral down-type squark $\tilde{d}_{Rk}$ and the RPV coupling $\lambda'_{21k}$, and perform a numerical study with Monte Carlo simulations.
As in the previous section we go through the tool chain: MadGraph 5 with a RPV-MSSM UFO model file\footnote{The model file can be found at \url{https://github.com/ilmonteux/RPVMSSM_UFO}.} and the same parton-level cuts, Pythia 6, FastJet 3, and Delphes 3 with the LHeC card.
Here we switch on only one single RPV coupling $\lambda'_{21k}$, for which the current (36 fb$^{-1}$) and projected (3 ab$^{-1}$) LHC bounds were recast in Ref.~\cite{Bansal:2018dge} from an ATLAS mono-lepton search \cite{Aaboud:2017efa}: $\lambda'_{21k}< 0.090\frac{m_{\tilde{d}_{Rk}}}{\text{1 TeV}}+0.014$ and $\lambda'_{21k}< 0.053\frac{m_{\tilde{d}_{Rk}}}{\text{1 TeV}}+0.029$.
The background process is $p \mu^- \to j \mu^-$ plus zero or one extra jet, for which we perform jet matching and merging.
Note that we ignore the subdominant effect from the interference terms.
For the cuts on the reconstructed events, we first select events with at least 1 jet.
We then specifically require that exactly one muon should be reconstructed.
We finally keep only events with the $p_T$ sum of the two leading jets, $p_T^{j1}+p_T^{j2}$, larger than certain values (for the events with exactly one jet we take $p_T^{j2}=0$).
We define the signal significance $S$ as $S=N_s/\sqrt{N_b}$ and determine the $95\%$ C.L. (confidence level) exclusion limits at $S=2$, where $N_{s/b}$ labels the signal/background event numbers.
The resulting limits on $\lambda'_{21k}$ as a function of $m_{\tilde{d}_{Rk}}$ for various $p_T$ sum thresholds are presented in Fig.~\ref{fig:rpv_nc}, which we overlap in red with the current LHC (solid) and future HL-LHC (dashed) bounds \cite{Bansal:2018dge}.
We find that increasing the lower threshold for the $p_T$ sum of the two leading jets allows to probe heavier $\tilde{d}_{Rk}$.
We conclude that $\mu p-1(2)$ may exclude values of $\lambda'_{21k}$ down to 0.02(0.01) for $m_{\tilde{d}_{Rk}}\sim\mathcal{O}(\text{TeV})$.
Compared to the HL-LHC, these $\mu p$ limits in $\lambda'_{21k}$ are stronger by up to more than one order of magnitude for $\tilde{d}_{Rk}$ light enough to be produced on shell.
The future hadron-hadron colliders such as the FCC-hh are expected to exclude SUSY squarks up to about 10 TeV \cite{stoppair_fcchh}. This is comparable to the $\mu p-1/2$ considered here.
As for future lepton colliders, $e^- e^+$ colliders are expected to perform much worse because of the relatively small center-of-mass energies, while multi-TeV muon colliders have recently been shown to possess huge potential for probing a similar theoretical scenario, i.e., the leptoquarks, possibly excluding the leptoquark mass at the order of 10 TeV \cite{Asadi:2021gah}.
We note that another possible signature with the same mediator and RPV coupling is the charged-current process $p \mu^-\to d \nu_\mu$.
However, we find that the exclusion limits in this scenario are similar to the NC results shown in Fig.~\ref{fig:rpv_nc}, and hence do not present the results here.

\section{Muon $g-2$ anomaly}\label{sec:muongm2}
One of the main drives for BSM physics has been the muon anomalous magnetic moment since about a decade ago.
With the latest world consensus on the SM computation of $a_\mu \equiv (g_\mu-2)/2$ \cite{Aoyama:2020ynm} combined with the experimental results published by the E821 collaboration at BNL \cite{Bennett:2006fi} and recently by the the Fermilab-based Muon $g-2$ experiment \cite{Abi:2021gix}, we are now faced with a discrepancy of $\sim 4.2 \sigma$ in $a_\mu$:
$
\Delta a_\mu = a_\mu^{\text{exp}} - a_\mu^{\text{SM}} = 251(59)\times 10^{-11}.~\footnote{There is a controversy arising from a new lattice calculation \cite{Borsanyi:2020mff} which shows a larger hadronic vacuum  polarization contribution, such that the total SM contribution to the muon $g-2$ is within $1\sigma$ of the experimental value.} \label{eqn:gm2_present}
$
One natural explanation could be weakly interacting NP appearing at the EW scale.
However, given the nonobservation of NP at the LHC so far, two other possibilities might be more relevant: (1) light NP below the GeV scale interacting feebly with the SM, and (2) much heavier NP (above the TeV scale) strongly coupled to the SM particles.
In this work, we consider the latter possibility.
If the NP scale, $\Lambda$, is much higher than the EW scale, i.e., $\Lambda\gg 1$ TeV, we can describe physics at energies much below $\Lambda$ with the framework of the Standard Model Effective Field Theory (SMEFT) \cite{Buchmuller:1985jz,Grzadkowski:2010es,Jenkins:2013wua,Alonso:2013hga}, of which the operators up to dim-6 relevant to $a_\mu$ 
are
\begin{eqnarray}
	\mathcal{L}\ni&& \frac{C^\mu_{eB}}{\Lambda^2}(\bar{\mu}_L \sigma^{\mu\nu}\mu_R) H B_{\mu\nu} + \frac{C^l_{eW}}{\Lambda^2}(\bar{\mu}_{L}\sigma^{\mu\nu} \mu_R)\sigma^i H W^i_{\mu\nu} \nonumber\\
	&&+\frac{C^\mu_T}{\Lambda^2}(\bar{\mu}^a_L \sigma_{\mu\nu} \mu_R)\epsilon_{ab}(\bar{Q}^b_L \sigma^{\mu\nu}u_R)+h.c. \label{eqn:SMEFTLagrangian}
\end{eqnarray}
The NP contributions to $a_\mu$ stem directly from the
operator $(\bar{\mu}_L \sigma_{\mu\nu}\mu_R)H F^{\mu\nu}$, which may be induced by the operators in Eq.~\eqref{eqn:SMEFTLagrangian} at tree or one-loop level.
Their corrections to $a_\mu$ can be expressed as follows:
\begin{eqnarray}
	\Delta a_\mu \simeq &&  \frac{4m_\mu v}{e \sqrt{2}\Lambda^2}\Big( C^\mu_{e\gamma} - \frac{3\alpha}{2\pi}\frac{c^2_W-s^2_W}{c_W s_W} C^\mu_{eZ} \text{log}\frac{\Lambda}{m_Z}  \Big)   \nonumber \\
	&&-\sum_{q=c,t}\frac{4 m_\mu m_q}{\pi^2}\frac{C_T^{\mu q}}{\Lambda^2}\text{log}\frac{\Lambda}{m_q}, \label{eqn:Deltaamu_gm2}
\end{eqnarray}
where $v=246$ GeV is the SM Higgs vacuum expectation value, $s_W$ and $c_W$ are sine and cosine of the Weinberg angle, and $C_{e\gamma}=c_W C_{eB} - s_W C_{eW}$ and $C_{eZ}=-s_W C_{eB} - c_W C_{eW}$.

At a $\mu p$ collider, only a limited set of these operators can be tested.
To start with, $C^\mu_{e\gamma}$ can be probed, either by considering the photon content inside the protons scattering an incoming muon, or by studying rare Higgs decays into a pair of muons and a photon.
We find the former possibility, suppressed by the parton distribution function of the photon in the protons, is insensitive to new physics that is sufficiently small to explaining the muon $g-2$ anomaly.
As for the rare Higgs decay, the decay branching ratio of the Higgs into $\mu^+ \mu^- \gamma$ should be at the order of $\sim 10^{-13}$ in order to test $\Delta a_\mu\sim 3\times 10^{-9}$ \cite{Buttazzo:2020eyl}.
However, at both $\mu p-1$ and $\mu p-2$ the production rates of the SM Higgs bosons are estimated to be roughly $10^5$ and $5\times 10^6$ (see Table~\ref{tab:VBFXS}), with 0.1 ab$^{-1}$ and 1 ab$^{-1}$ integrated luminosities, respectively, which 
are far from sufficient for probing a branching ratio of $10^{-13}$.
Consequently the only operator that could be confronted for $\Delta a_\mu\sim 3\times 10^{-9}$ at $\mu p-1$ and $\mu p-2$ is $C_{T}^{\mu c}$, with the parton-level process $\mu^- c \to \mu^- c$ and its $\bar{c}$ counterpart.
The corresponding background is $\mu^- p \to \mu^- j$.
Note that the unitarity constraint requires that $\Lambda \lesssim 10$ TeV for this operator. 

To explore the heavy NP in the muon $g-2$ at $\mu p-1$ and $\mu p-2$, we perform truth-level Monte Carlo simulations with the event generator MadGraph5 and the model package \texttt{SMEFTsim} \cite{Brivio:2017btx,Brivio:2020onw}, with the parton-level cuts $p_T^{j}>5$ GeV and $|\eta^{j}|<5.5$.
The computed cross sections for the signal and background processes are given in Table~\ref{tab:gm2_results}, assuming the contributions arise solely from the single SMEFT operator $C_{T}^{\mu c}$.
Therefrom we can easily obtain the signal and background event numbers, and hence the $2\sigma$ exclusion limits on $C^{\mu c}_T/\Lambda^2$.
To convert these limits into those on $|\Delta a_\mu|$, we take $\Lambda=10$ TeV for the logarithmic function in the last term of Eq.~\eqref{eqn:Deltaamu_gm2}, reaching $|\Delta a_\mu|=\frac{4 m_\mu m_c}{\pi^2}|\frac{C_T^{\mu c}}{\Lambda^2}|\text{log}\frac{10\text{ TeV}}{m_c}$.
The exclusion limits on $|\Delta a_\mu|$ are given in the last column of Table~\ref{tab:gm2_results}: $1.13\times 10^{-8}$ and $9.10\times 10^{-10}$.
\begin{table}
	\begin{center}
		\begin{tabular}{c|c|c|c}
			\hline
			& $\sigma^{\text{bgd}}~[\text{pb}]$ &$\sigma^{\text{sig}}\Big(\frac{\Lambda^2}{100\text{ TeV}^2C_{T}^{\mu c}}\Big)^2~[\text{pb}]$ & 	$|\Delta a_\mu |$\\
			\hline
			$\mu p-1$  &120 & $1.29\times 10^{-2}$   & $1.13\times 10^{-8}$\\
			$\mu p-2$  &30 & $3.15\times 10^{-1}$   & $9.10\times 10^{-10}$\\
			\hline 
		\end{tabular}
		\caption{
			Cross sections for the background ($\sigma^{\text{bgd}}$) and signal ($\sigma^{\text{sig}}$) events, as well as the integrated luminosities and the expected exclusion limits of $|\Delta a_\mu|$, at $\mu p-1$ and $\mu p-2$.
		}
		\label{tab:gm2_results}
	\end{center}
\end{table}
We conclude that in the limit of vanishing contributions from the other operators, the low-energy effects from the high-scale NP associated with the tensor operator $C^{\mu c}_T$ that would be small enough to explain the muon $g-2$ anomaly can be probed at $\mu p-2$.
In order to make $\mu p-1$ sensitive enough, further improvements on e.g., luminosity and search strategies, should be implemented. 

We note that since only $\Lambda \lesssim 10$ TeV is valid for the considered operator, it is necessary to check whether the typical hard-interaction CM energies for the signal are lower than 10 TeV.
We find that for $\mu p-1$, $\sim 100\%$ of the events have the invariant mass $m_{\mu c}$ of the outgoing muon and $c$ quark $m_{\mu c}< 5$ TeV, and for $\mu p-2$ it is about $\sim 70\%$ despite the much higher CM  energy.

\section{Conclusions}\label{sec:conclusions}
In this work we have proposed a muon-proton collider with two tentative configurations.
We performed numerical simulations to investigate the physics potential of $\mu p-1$ and $\mu p-2$ in both Higgs precision measurement and search for BSM physics.
Taking as benchmark physics cases the Higgs coupling to $b$-quarks, R-parity-violating MSSM, and heavy new physics in the muon $g-2$, we conclude that a multi-TeV muon-proton collider with $0.1-1$ ab$^{-1}$ integrated luminosities could show better performance than both current and future collider experiments.
Besides the physics scenarios studied here, we expect that this type of machine can also excel in other aspects of the SM precision measurements and BSM physics searches.
We believe this work could motivate more studies of TeV-scale muon-proton colliders in the high-energy physics community.

\begin{acknowledgments}
We would like to thank Florian Domingo and Jong Soo Kim for useful discussions.
This work was supported by MoST with grant nos. MoST-109-2811-M-007-509 and
107-2112-M-007-029-MY3.
\end{acknowledgments}

\bibliographystyle{h-physrev5}
\bibliography{bib}
\end{document}